\pdfoutput=1
\documentclass[aps,prd,reprint,groupedaddress,nofootinbib,preprintnumbers]{revtex4-1}

\usepackage{amssymb}
\usepackage{amsmath}
\usepackage[colorlinks=true,linkcolor=blue,citecolor=blue, urlcolor=blue]{hyperref}   

\usepackage{tikz}
\usepackage{graphicx}

\definecolor{uibred}{RGB}{167, 38, 47}
\def\Eq#1{Eq.~(\ref{#1})}
\def\Eqs#1{Eqs.~(\ref{#1})}
\def\eq#1{(\ref{#1})}
\def\app#1{Appendix~\ref{#1}}
\def\Fig#1{Fig.~\ref{#1}}

\def\Sec#1{Sec.~\ref{#1}}

\def\p{\mathbf{p}}

\def\k{\mathbf{k}}

\def\ptilde{\tilde{\p}}

\newcommand{\TId}{T_\text{id.}}
\newcommand{\Teff}{T^\text{eff}_\alpha}

\def\st{\begin{equation}}
\def\stp{\end{equation}}

\newcommand{\subfig}[2]{%
\begin{tikzpicture}%
\node[rectangle] (image) at (0,0) {#2};
\node[anchor=south west] (label) at (image.south west) {(#1)};
\end{tikzpicture}%
}
\begin{document}

\title{Chemical equilibration in weakly coupled QCD}

\author{Aleksi Kurkela}
\email[]{a.k@cern.ch}
\affiliation{Theoretical Physics Department, CERN, Geneva, Switzerland}
\affiliation{Faculty of Science and Technology, University of Stavanger, 
4036 
Stavanger, Norway}

\author{Aleksas Mazeliauskas}
\email[]{a.mazeliauskas@thphys.uni-heidelberg.de}
\affiliation{Institut f\"{u}r Theoretische Physik, Universit\"{a}t Heidelberg, 
69120 Heidelberg, Germany}
\preprint{CERN-TH-2018-239}

\date{\today}

\begin{abstract}

We study thermalization, hydrodynamization, and chemical equilibration in out-of-equilibrium Quark-Gluon Plasma starting from various initial conditions using QCD effective kinetic theory, valid at weak coupling. 
In non-expanding systems gauge bosons rapidly lose information of the initial state and achieve kinetic equilibrium among themselves, while fermions approach the equilibrium distribution only at later time.
In systems undergoing rapid longitudinal expansion, both gluons and quarks are kept away from equilibrium by the expansion, but the evolution is well described by fluid dynamics even before local thermal equilibrium is reached. 
For realistic couplings we determine the ordering between the separate hydrodynamization, chemical equilibration and thermalization time scales to be
 $\tau_\text{hydro}<\tau_\text{chem}<\tau_\text{therm}$.

\end{abstract} 
\maketitle

\section{Introduction}

How gauge theories pushed far from equilibrium thermalize is a central topic in the study of heavy-ion collisions~\cite{Busza:2018rrf}. To what extent the post-collisional debris created in the collision of two nuclei reaches local thermal equilibrium before the system cools down, determines how well a fluid dynamical description of the system is applicable. 

Our ability to perform first principles non-perturbative real-time calculations in QCD is limited by the infamous sign problem~\cite{Ding:2015ona}, and considerable efforts have been invested to understand thermalization and hydrodynamization in various approximations of the QCD. A prominent example is the $\mathcal{N} = 4$ Super Yang-Mills theory in the limit of a large number of colors and strong coupling, which has been studied extensively with holographic methods~\cite{Heller:2011ju,Heller:2015dha, Heller:2016rtz, Keegan:2015avk}. The holographic methodology can be applied only to a very limited set of gauge theories, and for generic theories---such as QCD---only weak coupling methods are available. So far the weak-coupling studies of  thermalization of far-from-equilibrium systems have been limited to either pure gauge or scalar theories, and studies in QCD have been restricted only to near equilibrium systems~\cite{Arnold:2003zc,York:2008rr}. Here, we extend the weak coupling treatment of \cite{Kurkela:2014tea, Kurkela:2015qoa} by including dynamical fermions, and study how far-from-equilibrium systems  approach equilibrium in a full leading order QCD description. 

Introducing new degrees of freedom to the system adds new structures. It has been argued~\cite{Biro:1981zi} that the off-equilibrium dynamics of quarks may be significantly slower than that of the gluons, owing partly to smaller group theoretic color factors, and partly to different spin statistics and Pauli blocking. 
It may be then that the
equilibration of quarks could be a bottleneck of thermalization as chemical equilibration may take place in a significantly longer time scale. In particular, in the weak
coupling picture of heavy-ion collisions, the initial state
in midrapidity is dominated by a large number of gluons
with only a few fermions. If the production of fermions is
delayed, this could have an impact on the fluid
dynamical modelling of heavy-ion collisions, since the equation of state of Quark-Gluon Plasma is different than that of plasma consisting of gluons only.
Furthermore, chemically equilibrated QGP is a standard explanation of the strangeness enhancement in nucleus-nucleus collisions~\cite{Cleymans:2006xj,Andronic:2008gu, Andronic:2017pug}, so understanding fermion production from first principles provides an important theoretical validation of this picture.

It has been been observed in several theories -- both weakly and strongly coupled -- that the hydrodynamical constitutive relations are approximately fulfilled in systems that have sizeable anisotropies, that is, hydrodynamization without thermalization \cite{Beuf:2009cx,Chesler:2010bi,Kurkela:2015qoa,Heller:2016rtz}. Upon including  quark degrees of freedom to the system we may ask the question, when does the chemical equilibration happen with respect to the hydrodynamization and thermalization times $\tau_{\rm hydro}$ and $\tau_{\rm therm}$?

The basic tool here is the set of Boltzmann equations that are
applicable in isotropic systems any time the typical occupancies are
smaller than $1/\alpha_s$.  Of course, many authors have already
considered the evolution of quark-gluon systems under Boltzmann
equations~\cite{Biro:1993qt,Baym:1984np,Geiger:1991nj,Zhang:1997ej,
Borchers:2000wf,Xu:2004mz,Blaizot:2014jna,Huang:2013lia,Scardina:2014gxa}. 
What sets our study apart from these works is that we use the effective Boltzmann equations derived by Arnold, Moore, and Yaffe
\cite{Arnold:2002zm}, which account for all processes needed for a description that is accurate to leading order in $\alpha_s$. These 
processes include in-medium screening effects~\cite{Braaten:1989mz}, and Landau-Pomeranchuk-Migdal~\cite{Landau:1953gr, Landau:1953um,Migdal:1956tc,Migdal:1955nv} corrected splitting processes in
far-from-equilibrium but isotropic systems\footnote{We note that there are also other non-perturbative (in a sense that they involve an infinite number of diagrams), but weak-coupling (in the sense that they can be expanded in a series in the coupling constant) processes which
do not contribute to our leading order calculation. These processes include for example the sphaleron transitions, see \cite{Mace:2016svc}.}.  To do so we extend the previously developed setup of Refs.~\cite{Kurkela:2014tea, Kurkela:2015qoa,Keegan:2015avk, Keegan:2016cpi} to include quark degrees of freedom and we briefly summarize the key elements of the description in \Sec{sec:EKT}.

In the following  we will consider $N_c=3$ QCD plasma with $N_f =3$ flavours of massless quarks in  different out-of-equilibrium 
conditions, with and without expansion.  While there is only one thermal equilibrium, there are many ways how a
system can be out of equilibrium. In \Sec{sec:IIIA} we will first study in detail a particularly simple non-expanding system where  
quarks are absent and gluons are in kinetic equilibrium among themselves. We discuss the processes that produce fermions and subsequently lead to chemical equilibration. In \Sec{sec:IIIB} we then move on to discuss non-expanding systems which are initialized with an overoccupied gauge boson distribution. {While we are unaware of a physical system where QCD would 
be found in these conditions, cosmological reheating may result in system of overoccupied
gauge bosons (see e.g. \cite{Figueroa:2015rqa}).} Following the time evolution of the overoccupied system, we see that 
due to the slower dynamics of  fermions, gauge bosons  reach kinetic equilibrium before 
the chemical equilibration. Once the gauge bosons have reached kinetic equilibrium the evolution
proceeds as in our first example. 
Finally, in \Sec{sec:exp} we  turn to a system of overoccupied gluons undergoing boost-invariant longitudinal expansion. This is the expected initial condition in heavy-ion collisions in the asymptotic weak coupling limit~\cite{Lappi:2006fp, Gelis:2010nm}. For moderate values of the coupling constant $\alpha_s\sim 0.3$ we observe a rapid memory loss of initial conditions and the chemical and hydrodynamical equilibrium is approached along a universal curve.
Finally, we conclude with the discussion of the separate equilibration time scales in \Sec{sec:conclusions}.

\section{Effective kinetic theory\label{sec:EKT}}
The effective kinetic theory that we use to describe thermalization is the 
Effective Kinetic Theory (EKT) of Arnold, Moore and Yaffe~\cite{Arnold:2002zm}, 
which is leading order accurate in the QCD coupling constant $\lambda= g^2 N_c=  4\pi \alpha_s N_c$ in the combined limit
of weak coupling $(\lambda \rightarrow 0)$ and nonperturbative occupancies
$(\lambda f \rightarrow 0)$ for modes whose momenta are larger than the 
thermal screening scale in the nonequilibrium system $p^2 \gg m^2 \sim \lambda\int d^3 \p f(\p)/p$. 

At leading order in the coupling constant, the EKT describes the time evolution of color/spin averaged distribution function $f_s$ 
with an effective $2\leftrightarrow 2$ scattering and a $1\leftrightarrow 2$
effective splitting terms. The resulting Boltzmann equation for homogeneous non-expanding system is
 \begin{align}
\partial_t f_s(\p,t) &= -\mathcal{C}^s_{2\leftrightarrow 2}[f](\p)-\mathcal{C}^s_{1\leftrightarrow 2}[f](\p)
\label{eq:bolz}
 \end{align}
 with massless dispersion relation, $p^0 = |\p |=p$.
The index $s$ refers to different particle species in the theory. Expanding upon previous implementations of pure gauge theories in Refs.~\cite{Kurkela:2014tea, Kurkela:2015qoa,Keegan:2015avk, Keegan:2016cpi} to QCD,  $s$ now stands for gluons and $2N_f$ massless fermions (with quarks and anti-quarks counted separately)\footnote{In this work we consider plasma at zero chemical potential with quark and anti-quark distributions being equal.}.  
 The symmetrized $2\leftrightarrow 2$ collision terms in the right hand side of Eq.~(\ref{eq:bolz}) reads
\begin{widetext}
\begin{align}
\label{eq:2to2}
\mathcal{C}^s_{2\leftrightarrow 2}[f](\ptilde) &= \frac{1}{2}
\frac{1}{\nu_s} \frac{1}{4}
 \sum_{abcd} \int_{\p\k\p'\k'} 
 |\mathcal{M}^{ab}_{cd} 
 |^2(2\pi)^4 
 \delta^{(4)}(p^\mu+k^\mu-p'^\mu-k'^\mu)\\
&\times \{ (f^a_\p f^b_\k (1\pm f^c_{\p'})(1\pm f^d_{\k'}))-(f^c_{\p'} 
f^d_{\k'} (1\pm f^a_{\p})(1 \pm f^b_{\k})) \}\nonumber\\
&\times (2\pi)^3
\left[\delta^{(3)}(\ptilde-\p)\delta_{as}+\delta^{(3)}(\ptilde-\k)\delta_{bs} 
-\delta^{(3)}(\ptilde-\p')\delta_{cs}-\delta^{(3)}(\ptilde-\k')\delta_{ds}\right],\nonumber
\end{align}
\end{widetext}
where $|\mathcal{M}^{ab}_{cd}|^2$ is a $2\leftrightarrow2$ scattering amplitude-squared summed over all degrees of freedom of the external legs ($\nu_q=2N_c$ 
for quarks and 
$\nu_g=2 (N_c^2-1)$ for gluons), $\sum_{abcd}$ is a sum over all particle and antiparticle species, and  $\int_\p = \frac{d^3 \p}{2p (2\pi)^3}$ is a shorthand notation for  Lorentz invariant momentum integral. The second line is the usual phase-space loss and gain terms, while the Kronecker and Dirac delta functions in the last line accounts for the possibility of particle $s$ to be on any of the four external lines. Finally the numerical prefactors in front of the integral correct the double counting of identical processes.

The effective matrix elements $|\mathcal{M}^{ab}_{cd}|^2$  in \Eq{eq:2to2}  are for most kinematics the normal tree-level vacuum matrix element (see Table II in Ref.~\cite{Arnold:2002zm}). For soft small angle scatterings with energy transfer $\omega \ll p,k$, the tree-level Coulomb and Compton scatterings are infrared divergent elevating a set of diagrams with an arbitrary number of loops to the same magnitude as the tree-level diagrams. These effects become important at the in-medium screening scale $p \sim m_g, m_q$.
Here the in-medium effective masses of gluon and quarks are given, respectively, by  
\begin{align}
&m_{g}^2 =  2 g^2 \int_\p\big[ 2C_A f_g(\p)+ 2N_f C_F \frac{\nu_q}{\nu_g} (f_q(\p)+f_{\bar q}(\p))\big],\\
&m_{q}^2 =  2 g^2  \int_\p\big[2C_Ff_g(\p)+C_F(f_q(\p)+f_{\bar q}(\p))\big].
\end{align}
For momentum transfer of this order the dispersion of the internal line in the computation of $|\mathcal{M}^{ab}_{cd}|^2$ gets an $\mathcal{O}(1)$ correction from the in-medium physics. 
Therefore, for the problematic soft scattering we replace the matrix element with that computed in the Hard Thermal Loop (HTL) approximation that self-consistently treats the medium interaction correctly to leading order. We perform this substitution by removing the infrared divergent small angle approximation from the full matrix element and replace it with the small angle approximation of the full HTL rate~\cite{York:2014wja}. Specifically for a soft gluon or fermion exchange  with the momentum 
transfer $q = |\p' - \p|$  in the $t$-channel, the divergent term
${(u-s)}/{t}\sim {1}/{q^2}$
is replaced by IR regulated term
\begin{equation}
\frac{u-s}{t}\rightarrow\frac{u-s}{t}\frac{q^2}{q^2+\xi_s^2 m_s^2},\label{eq:reg}
\end{equation}
where $\xi_g = e^{5/6}/2 $ and $\xi_q=e/2$ are fixed such 
that the matrix element 
reproduces the full HTL results for
drag and momentum diffusion properties of soft gluon scattering~\cite{York:2014wja} and gluon to quark conversion $gg\rightarrow q\bar q$~\cite{Ghiglieri:2015ala,teaney} at leading order for isotropic distributions.

\def\n{\hat{\mathbf{n}}}
\def\h{{\mathbf{h}}}
\def\q{{\mathbf{q}}}
While the soft $\omega \sim m_g$ scatterings do not appreciably change the momentum state of the particle, 
they may bring the particle slightly off shell and make it kinematically possible for the particle to decay through
nearly collinear splitting. This makes the effective $1\leftrightarrow 2$ matrix element a leading order effect. It is included as $\mathcal{C}^{1\leftrightarrow 2}[f](\tilde \p)$ on the right hand side of the Boltzmann equation \Eq{eq:bolz} and explicitly
\begin{align}
\label{1to2}
&\mathcal{C}^s_{1\leftrightarrow 2}[f](\tilde \p)=\\
&= \frac{1}{2}\frac{1}{
\nu_s}\sum_{abc}\int_0^{\infty} dp dp' dk' \,  4\pi\gamma^{a}_{bc}(p;p',k')\delta(p-p' 
-k')\nonumber \\
&
\times
\big\{f_{p\n}^a[1\pm f^b_{p'\n}][1\pm f^c_{k'\n}] - f^b_{p'\n}f^c_{k'\n}[1\pm f^a_{p\n}]\Big\}\nonumber\\
&\nonumber \times \frac{(2\pi)^3}{4\pi\tilde p^2}[\delta(\tilde p-p)\delta_{as}-\delta(\tilde p-p')\delta_{bs}-\delta(\tilde p-k')\delta_{cs}],
\end{align}
where the unit vector $\n=\tilde \p/|\tilde \p|$ defines the splitting direction and $\gamma^a_{bc}(p; p', k')$ is the effective collinear splitting rate including Landau-Pomeranchuk-Migdal~\cite{Landau:1953gr, Landau:1953um,Migdal:1956tc,Migdal:1955nv} suppression of
collinear radiation.
Factoring out the kinematic splitting functions the rates
\begin{align}
\gamma^g_{gg}(p;p',k')=&\frac{p^4 + p'^4 + k'^4}{p^3 p'^3 k'^3} \mathcal{F}_g(p; 
p',k'),   \\
\gamma^q_{qg}(p;p',k')=&\frac{p^2 + p'^2}{p^2 p'^2 k'^3} \mathcal{F}_q(p; 
p',k'),  \\
\gamma^g_{q\bar{q}}(p;p',k')=&\gamma^q_{qg}(k'; 
-p',p)  
\end{align}
are given by an effective vertex resuming an infinite number of possible soft interactions with the medium~\cite{Arnold:2002zm}. It is found by solving the following integral equation
\begin{align}
 2 \h =& i \delta E(\h) \mathbf{F}_s(\h) + g^2 T_* \int 
 \frac{d^2 \q_\perp}{(2\pi)^2} \mathcal A(\q_\perp)  \label{diffeq}\\
 \times  &\Bigg\{ 
\frac{1}{2} \left(C_s  +C_s - C_A\right)\left[\mathbf{F}_s(\h)
                                    - \mathbf{F}_s(\h-k' \q_\perp) 
                                  \right] \nonumber \\
&   +                        \frac{1}{2} \left(C_s  +C_A - C_s\right)\left[ 
\mathbf{F}_s(\h)  - \mathbf{F}_s(\h-p' \q_\perp)\right] 
\nonumber \\
&  +                         \frac{1}{2} \left(C_A  +C_s - C_s\right) \left[ 
\mathbf{F}_s(\h)  - \mathbf{F}_s(\h+p 
\q_\perp)\right]\Bigg\}\nonumber
\end{align}
and $ \mathcal{F}_s(p; p',k')$ is defined as
\begin{align}
 \mathcal{F}_s(p; p',k') =& \frac{\nu_s C_s g^2}{8(2\pi)^4}\int \frac{d^2 
 h}{(2\pi)^2} 2 \h \cdot \text{Re}\, \mathbf{F}_s(\h; p, p', k') \label{curlF}.
\end{align}
In this work the strength of soft momentum background fluctuations $\mathcal A(\q_\perp)$ are treated using  an isotropic screening approximation~\cite{Aurenche:2002pd}
\begin{equation}
\mathcal A(\q_\perp)= \frac{1}{\q_\perp^2}- 
 \frac{1}{\q_\perp^2+2m_g^2},
\end{equation}
 the energy difference $\delta E$ is defined as 
\begin{align}
\delta E(\h; p,p',k')&\equiv \frac{m_{g}^2}{2 k'}+\frac{m_{s}^2}{2 p'} 
- \frac{m_{s}^2}{2 p}+ \frac{\h^2}{2 p k' p'},
\end{align}
and the effective temperature $T_*$ is given by
\begin{equation}
T_* \equiv \frac{1}{\nu_g m_g^2}\sum_s \nu_s g^2 C_s\int 
\frac{d^3p}{(2\pi)^3} f_s(\p)(1\pm f_s(\p)).
\end{equation}
Instead of solving \Eq{diffeq} directly, the required two dimensional integral \Eq{curlF} is expressed as the value at the origin of the Fourier transformed function $\tilde {\mathbf{F}}_s$, which solves a Fourier transformed  \Eq{diffeq}. We solve it using the basis function method \cite{Ghiglieri:2014kma} and parametrize the solution for the Monte-Carlo sampling of the collision kernel in \Eq{1to2}. 
For the distribution functions we use a previously developed discretization scheme that 
does not introduce additional discretization errors for energy or particle number densities~\cite{York:2014wja,Kurkela:2015qoa,Keegan:2015avk}.
The distribution functions are discretized in spherical polar coordinates on logarithmic momentum grid with typical momentum range $p_\text{max}/p_\text{min}=1500$ and $N_p=100$. For anisotropically expanding systems the longitudinal momentum fraction $\cos\theta=p_z/p$ is discretized on a uniform 
grid with typical value of $N_\theta=200$. We considered azimuthally symmetric distributions.
The collision integrals were calculated by Monte Carlo sampling of the phase space with importance sampling~\cite{Keegan:2015avk}.
At each time step the $2\leftrightarrow 2$ collision integral \Eq{eq:2to2}  was calculated using  $N_{2\leftrightarrow 2}N_pN_\theta$  randomly generated vector quadruplets $\p,\k,\p',\k'$ satisfying the momentum and energy conservation, where $N_{2\leftrightarrow 2}=50,100$. For $1\leftrightarrow 2$ collision integral \Eq{1to2} we used $N_{1\leftrightarrow2}N_p$ samples of momentum  $p,p',k'=p-p'$ combinations, which were reused for each angular direction (here $N_{1\leftrightarrow2}=50,100$).

\section{Chemical equilibration in isotropic non-expanding systems\label{sec:results}}

\subsection{Kinetically equilibrated initial conditions\label{sec:IIIA}}
 In order to gain intuition to the far-from-equilibrium dynamics, we start with a particularly simple system where 
gauge bosons and fermions are initialized at time $t= 0$ with thermal distributions
\begin{align}
f^s_\text{eq}(p) = \frac{1}{e^{p/T_{s,{\rm init.}} }\pm 1},
\end{align}
but with different initial temperatures $T_{g, \rm init.}\neq T_{q, \rm init.}$. In such a situation we say that quarks and gluons are in \emph{kinetic equilibrium} among themselves, but not in thermal equilibrium with each other. 
The system will relax into state in which both the quarks and the gluons are in equilibrium with the same temperature $T_\text{final}$.
 The energy conservation dictates that the final temperature will be given by
\begin{equation}
\nu_gT^4_{g,\text{init.}} + 2N_f\nu_q \tfrac{7}{8}T^4_{q,\text{init.}}=(\nu_g + 2N_f\nu_q \tfrac{7}{8})T_\text{final}^4.\label{eq:Tfinal}
\end{equation}

\begin{figure}
\centering
\includegraphics[width=0.9\linewidth]{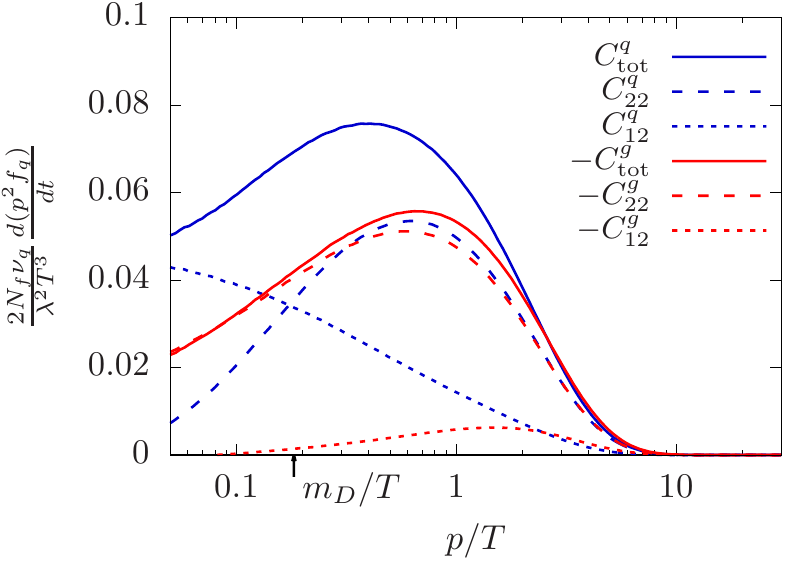}
\caption{The rate of change of total fermion number density (blue lines) for a thermal gluon system with no initial quark density and $\lambda=0.1$. The dashed and dotted lines show the contributions to the rate from elastic $2\leftrightarrow2$ and inelastic $1\leftrightarrow 2$ processes, while the solid lines are the sum of the two. The corresponding changes in the gluon number density (multiplied by $-1$) are shown by red lines.\
}
\label{fig:plotccollnumberl0}
\end{figure}

If we start with pure gluon initial state, \emph{i.e.}, $T_{q,\text{init.}}=0$, the fermion number is initially zero. In order to reach chemical equilibrium, the fermion number has to be subsequently generated by pair production either through medium induced $g\rightarrow q\bar q$-splitting processes or alternatively through $gg\rightarrow q \bar q$ conversions. Multiplying the Boltzmann equation \Eq{eq:bolz} by  $2N_f \nu_q p^2/{\lambda^2 T^3}$ for quarks, we obtain the equation for the rate of change of fermion number (per momentum)
 \begin{align}
 \frac{2N_f\nu_q}{\lambda^2 T^3}\partial_t [p^2 f_q(\p,t)]&= C_{22}^q+C_{12}^q
 \end{align}
and similarly for gluons.
 In \Fig{fig:plotccollnumberl0} we show the rates $C_{22}^s$ and $C_{12}^s$ separately and as a sum for coupling constant $\lambda=0.1$ and temperature $T=T_{g,\text{init}}$. We see that the $2\leftrightarrow 2$ processes dominate  fermion production around $p\sim T$ (blue dashed line), while for $p\sim m_D = \sqrt{2} m_g$ the splitting processes become roughly equally important (blue dotted line).   
The changes in the gluon distribution mirror the fermionic ones.
 We see that $gg\rightarrow q \bar q$ conversion reduces the number of gluons at the same momentum scale $p\sim T$  where fermions are created (blue and red dashed lines), while soft collinear radiation from  $p\sim T$ gluons (red dotted line) produces soft fermions. Importantly, the resulting fermion spectrum is non-thermal.

\begin{figure}
\subfig{a}{\includegraphics[width=0.9\linewidth]{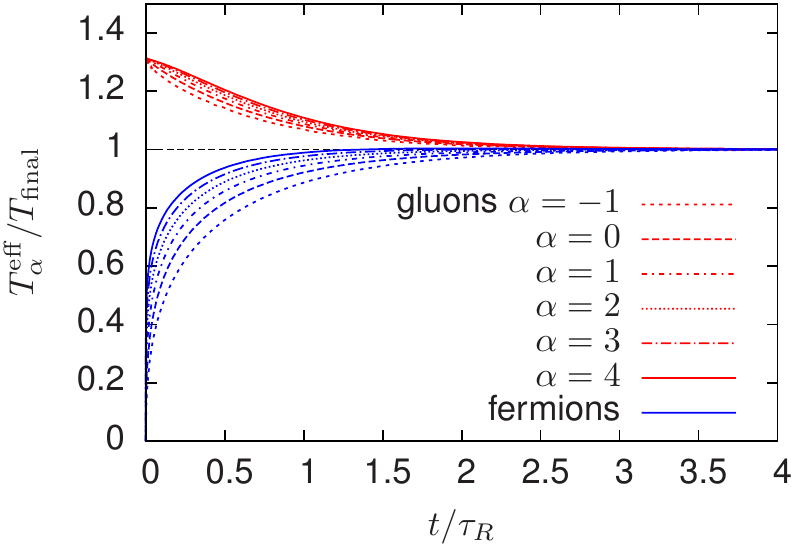}}
\caption{
The various effective temperatures (see. \Eq{eq:Teff}) as a function of time for a non-expanding system with initial conditions of thermal gluons and no fermions. The time axis is scaled by $\tau_R=4\pi \eta/s/T_\text{final}$, where $\eta/s\approx 1900$ for  $\lambda=0.1$.
 The red lines are the gluonic effective temperatures; the blue lines are the effective temperatures of the fermionic sector with $T^\text{eff}_4>T^\text{eff}_3> \ldots> T^\text{eff}_{-1}$.
\label{fig:Talpha1}}
\end{figure}

We now turn to study how the above system evolves toward equilibrium as a function time. For the same $\lambda = 0.1$ as above, Fig.~\ref{fig:Talpha1} displays the
time evolution of several different \emph{effective temperatures} $T^{\rm eff}_{g/q,\alpha}$, defined through the $\alpha$th moments of the fermion ($F=1$) and boson ($F=0$) distribution functions
\begin{equation}
T_{g/q,\alpha}^{\rm eff} = \left[ \mathcal{N}_{\alpha,F}  \int \frac{d^3 p}{(2\pi)^3}  p^\alpha f(p)\right]^{\frac{1}{\alpha+3}}.
\label{eq:Teff}
\end{equation}
The effective temperatures are normalised such that when the system is in thermal equilibrium all $T_{g/q,\alpha}^{\rm eff}$ (for all $\alpha$) are equal to the equilibrium temperature $T_{g/q,\alpha}^{\rm eff}=T_\text{final}$, that is
$$
\mathcal{N}_{\alpha, F} = [1-2^{-\alpha-2}]^{-F} \frac{2\pi^2 }{\Gamma(\alpha+3)\zeta(\alpha+3)}.
$$
Lower $\alpha$ values are more sensitive to the infrared of the distribution, whereas larger $\alpha$ values describe the UV part of the distribution function. In particular, $T^\text{eff}_1\propto \sqrt[4]{e}$ corresponds to the temperature of a fictitious thermally equilibrated system with the same energy density.

We display the effective temperatures as a function of \emph{relaxation time} of the final thermalized system
\begin{equation}
\tau_R = \frac{4\pi \eta/s}{T_{\rm final}},
\end{equation}
where $\eta/s$ is the specific shear viscosity, whose relation to $\lambda$ is discussed in \app{sec:etas}. This relaxation time
is parametrically of order of the transport mean free path of the thermalized system $\tau_R \sim (\lambda^2 T_{\rm final})^{-1}$, 
but it has been observed in \cite{Kurkela:2015qoa,Keegan:2015avk,Kurkela:2018vqr,Kurkela:2018wud} that expressing the coupling constant $\lambda$ in terms of the specific viscosity accounts
for large numerical corrections that go beyond the parametric expression and leads to better scaling behaviour for different values of $\lambda$.

We first note that the effective temperatures of gluons (red lines in Fig.~\ref{fig:Talpha1}) decrease while those of the quarks (blue lines) increase. During this evolution the gluon effective temperatures approximately overlap signifying that gluons remain close to the kinetic equilibrium through the whole evolution. In contrast to gluons---and consistent with the non-equilibrium spectrum of the fermion production rate in Fig.~\ref{fig:plotccollnumberl0}---the fermion effective temperatures differ until full chemical equilibration is achieved. During this evolution the fermion spectrum is harder than that in kinetic equilibrium as seen from the ordering of the effective temperatures $T_4^\text{eff} > T_3^\text{eff} > \ldots >  T_{-1}^\text{eff}$.  

\begin{figure}
\centering
\includegraphics[width=0.9\linewidth]{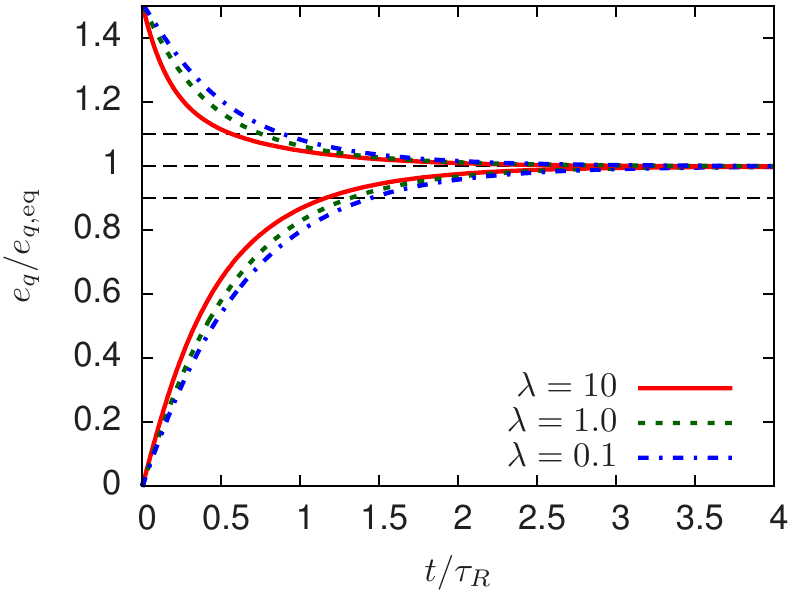}
\caption{Chemical equilibration of fermion energy density (as a function of equilibrium energy density) for different coupling constants $\lambda=10,1.0,0.1$, which corresponds to effective $\eta/s=1,35,1900$. Two cases shown: with zero initial fermion energy $e_q\approx 0$ and with zero initial gluon energy $e_g\approx 0$.
}
\label{fig:plotethgfnonexpx}
\end{figure}

In chemical equilibrium $N_f=3$ fermions constitute $e_{q,\text{eq}}/e_{q,\text{total}}\approx 0.66$  of the total equilibrium energy density (see \Eq{eq:Tfinal}).
It can be seen in \Fig{fig:plotethgfnonexpx} that by the time $t_\text{chem} \approx 1.4 \tau_R$ for $\lambda=0.1$
the fermion energy density $e_q$ has reached 90$\%$ of its equilibrium value $e_{q,{\rm final}}$, which
we take as our somewhat arbitrary definition of the chemical equilibration time, i.e.
\begin{equation} 
\frac{e_q(t_\text{chem})}{e_{q,{\rm final}}} = \left(\frac{T^{\rm eff}_{1,q}(t_\text{chem})}{T_{\rm final}}\right)^4= 0.9 \label{eq:chem}.
\end{equation}

To study the coupling dependence of the chemical equilibration, we repeat the above calculation with the same initial conditions, but with several different values of $\lambda = 10,1.0,0.1$, corresponding to $\eta/s \approx 1,35, 1900$, respectively. Figure \ref{fig:plotethgfnonexpx} shows the time-evolution of fermion energy fraction for 
these different values of 't Hooft couplings, for which the chemical equilibration time varies by three orders of magnitude. However, as is seen from the figure, the functional forms of the time-evolutions of the energy densities follow closely a common form when described in terms of the relaxation time $\tau_R$, with the system reaching chemical equilibrium around 
\begin{equation}
t_\text{chem}/\tau_R \sim 1.1-1.4
\end{equation}
for all studied values of $\lambda$. 

We  note however that for moderate values $\lambda \gtrsim 10$, we expect substantial next-to-leading  order corrections to  transport properties of QGP~\cite{Ghiglieri:2018dib}.
 Nevertheless, while the kinetic theory suffers from large systematic uncertainties for these values of $\lambda$, the fact that the kinetic theory calculation itself does not fail catastrophically may allow one to extrapolate the results for semi-quantitative order of magnitude estimates even at larger values of $\lambda$ using $\tau_R$ scaling, where the dependence on the coupling constant only enters through the specific shear viscosity $\eta/s$.  Note that other transport coefficients in units of $\eta/s$, e.g. $\tau_\pi/(\eta/(sT))$, are much less sensitive to the changes of the coupling constant~\cite{Ghiglieri:2018dgf}.

We note in passing that similar qualitative features of the equilibration are seen by varying the initial starting gluon and fermion temperatures. In particular, the gluons remain in approximate kinetic equilibrium throughout evolution even if one starts with fermion dominated initial conditions (data not shown). In this case the fermion energy fraction approaches equilibrium ratio from above (see \Fig{fig:plotethgfnonexpx}) and the chemical equilibration takes place at $t_\text{chem}/\tau_R\sim 0.5-0.9$, where now $e_q(t_\text{chem})/e_{q,\text{final}}=1.1$ in this case.

\subsection{Overoccupied initial conditions\label{sec:IIIB}}

\begin{figure}
\centering
\includegraphics[width=0.9\linewidth]{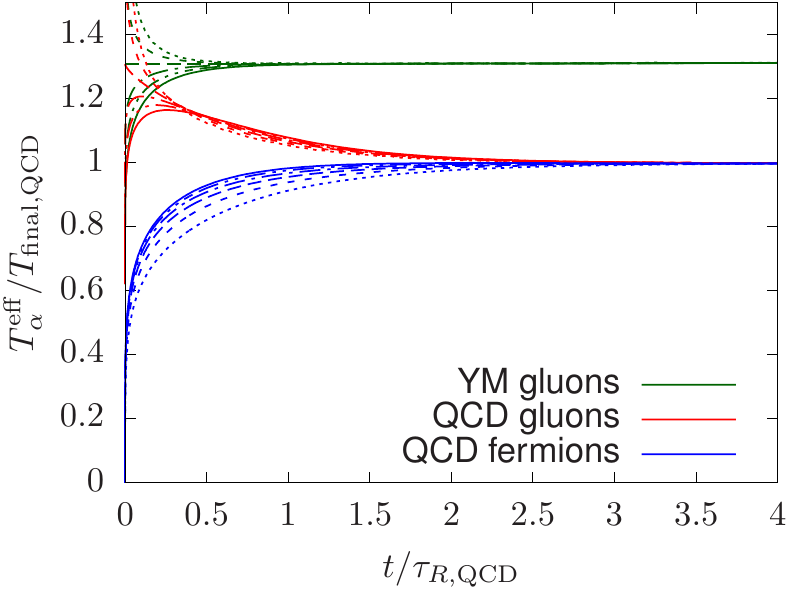}
\caption{ Chemical equilibration for isotropic over-occupied gluon state with no initial fermions present.
The effective temperature $T_\alpha$ of gluon (red  lines) and fermion (blue lines) distribution functions are shown for $\lambda=0.1$. Note that for fermions $T^\text{eff}_4>T^\text{eff}_3> \ldots> T^\text{eff}_{-1}$, but over-occupied gluons start with inverted ordering.}
\label{fig:plotteffprlgfnonexp10}
\end{figure}

We now turn to a system starting with an overoccupied distribution of gluons and no fermions. This system
has initially too many gluons, $f_g \gg 1$, with too soft momenta compared to a thermal ensemble with the same
energy density, that is $T^{\rm eff}_2 \ll T^{\rm eff}_1 \ll T^{\rm eff}_0$. As the fermions cannot be overoccupied such a system is necessarily dominated by the gluons in the
early stages. As discussed in detail in \cite{Berges:2008mr,Berges:2012ev,Schlichting:2012es,Berges:2013fga, Kurkela:2011ti,York:2014wja, Kurkela:2014tea,Kurkela:2012hp}, overoccupied gluonic systems thermalize via a self-similar cascade
which carries the energy and particle number from the infrared to the ultraviolet via elastic and inelastic scattering.
As long as the system is parametrically overoccupied $f_g\gg 1$, the cascade is self-similar, that is, the gluon distribution
function at a given time $t$ is given by an approximately stationary scaling function $\tilde f$  which is insensitive to the initial
conditions
\begin{equation}
f_g(p) = (Q t)^{-4/7} \tilde f( p/p_\text{max}), \quad  p_\text{max} = Q (Qt)^{1/7},\label{eq:sc}
\end{equation}
where $Q^4 = 2\pi^2 \lambda \int_\p p f_ g (p)$. 
This scaling form is reached in a time that is proportional to the scattering rate of
the initial condition~\cite{Kurkela:2014tea}, which is parametrically faster than the thermalization time for a parametrically overoccupied
system. The approach to the scaling form is discussed in detail in \cite{Kurkela:2012hp}.
Once the typical momentum of the cascade  $p_{\rm max} $ reaches the thermal scale $T_\text{final}$, the system equilibrates.
How the presence of fermions may affect the cascade has been studied using (semi-)classical Yang-Mills simulations in the regime where the gluons are still highly overoccupied and $p_{\rm max} \ll T_{\rm final}$~\cite{Gelfand:2016prm}. We now answer the question what happens when $p_{\rm max} \sim T$. 

\begin{figure}
\centering
\includegraphics[width=0.9\linewidth]{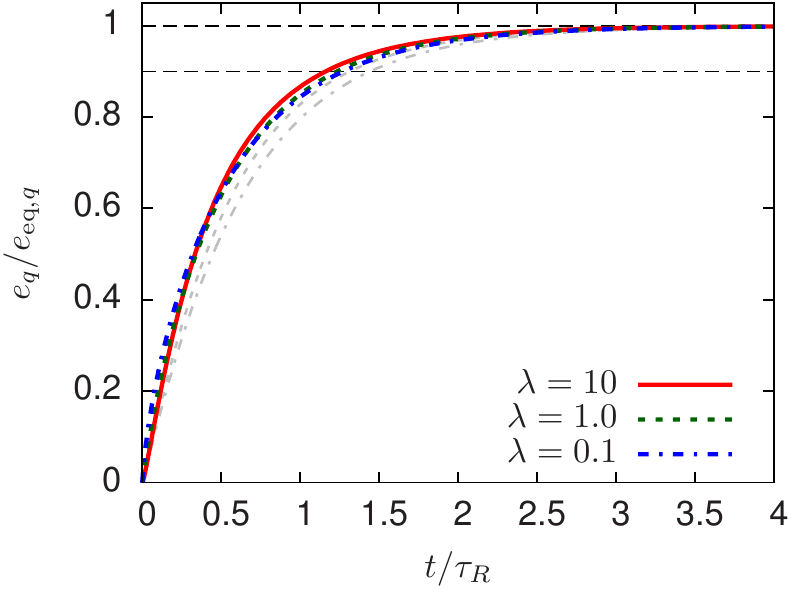}
\caption{Chemical equilibration of fermion energy density (as a function of equilibrium energy density) for different coupling constants $\lambda=10,1.0,0.1$, which correspond to effective $\eta/s=1,35,1900$ for the system initialized with over-occupied gluon density (colored lines) and thermal initial conditions (gray lines).}
\label{fig:prl_e_nonexp}
\end{figure}

The green lines in \Fig{fig:plotteffprlgfnonexp10} show the time evolution of a pure $N_c=3$ gauge theory with over-occupied gluon initial conditions\footnote{The initial condition of the simulation is given by \Eq{eq:focc} with $\lambda=0.1$ and $\xi=1$. However as the system very quickly reaches the self-similar scaling form, the details of the initial condition matter only at very early times $t \ll \tau_R$.}.
For the effective temperatures, this scaling form \Eq{eq:sc} corresponds to a power-law behaviour 
\begin{equation}
T^\text{eff}_\alpha \propto t^{\frac{1}{7}\frac{\alpha-1}{\alpha+3}}.
\end{equation}
 If the system
is parametrically overoccupied, the effective temperatures are also parametrically separated $T^\text{eff}_\alpha \gg T^\text{eff}_{\alpha+1}$, but when the
system thermalizes around $t \approx 0.3 \,\tau_R$, the effective temperatures again collapse.
\Fig{fig:plotteffprlgfnonexp10} also shows the time evolution of a QCD plasma ($N_c = 3$, $N_f = 3$) with the same initial condition of overoccupied gluons with no quarks present. The time evolution differs now from the pure glue system as the fermion number is dynamically generated. At early times the quark effective temperatures are small and gluonic temperatures approximately follow the classical
power-laws. The cascade ends and the gluonic sector reaches a kinetic equilibrium among themselves again around 
\begin{equation} 
t \sim t_\text{kinetic}^g \approx 0.2 \tau_R,
\end{equation}
where we have defined the \emph{kinetic equilibration time}---in analogy with the chemical equilibration time---by demanding
\begin{equation}
\left(
\frac{T^g_0(t_\text{kinetic})}{T^g_1(t_\text{kinetic})}
\right)^4 = 0.9
\end{equation}
This timescale is significantly faster than the timescale
for chemical equilibration $t_\text{chem}$. Indeed at the time the kinetic equilibrium among gluons is reached there are only a few quarks present and the state of the system is approximately the same as in the case of thermal initial
conditions studied in the previous section. From this point on, the chemical equilibrium follows the same pattern which was described in the previous section. To emphasize this point, in \Fig{fig:prl_e_nonexp} we show the fermion energy fraction for over-occupied initial conditions (colored lines)
on top of the thermal initial conditions (gray lines), which were shown in \Fig{fig:plotethgfnonexpx}.

{

\section{Chemical equilibration in expanding systems\label{sec:exp}}

In the weak coupling description of heavy-ion collisions the state of the system right after the initial particle production is given by an overoccupied distribution of gluons \cite{Lappi:2006fp, Gelis:2010nm}. The novel feature compared to the previous section is that the geometry of the collision system is such that the overoccupied matter is undergoing a rapid, approximately boost invariant longitudinal expansion. Such system has been studied in detail in pure gauge theory \cite{Kurkela:2015qoa,Keegan:2016cpi} and it forms the link between the initial condition and the hydrodynamic stage in phenomenological multi-stage simulations of heavy-ion collisions \cite{Kurkela:2018vqr, Kurkela:2018wud}. Here, we study the expanding plasma in full QCD.

Assuming boost invariant form of the distribution function, the Boltzmann equation can be written in the form~\cite{Baym:1984np}
 \begin{align}
\partial_\tau f_s(\p,\tau) - \frac{p^z}{\tau }\partial_{p^z}f_s(\p)
 &= -\mathcal{C}^s_{2\leftrightarrow 2}[f](\p)-\mathcal{C}^s_{1\leftrightarrow 2}[f](\p),
 \end{align}
where $\tau$ is the Bjorken time $\tau=\sqrt{t^2-z^2}$. The expansion redshifts the distribution in the $p^z$ direction making it more anisotropic along the longitudinal momentum, while $2\leftrightarrow 2$ scatterings act to isotropize the system~\cite{Baier:2000sb}. While anisotropic systems could suffer from the presence of unstable plasma modes~\cite{Mrowczynski:1988dz, Mrowczynski:1993qm,Mrowczynski:2000ed} affecting the kinetic dynamics~\cite{Kurkela:2011ub,Kurkela:2011ti}, the detailed 3+1D classical-statistical Yang-Mills simulations~\cite{Berges:2013fga,Berges:2013eia} found that late time evolution of anisotropic systems is in agreement with kinetic theory expectations neglecting plasma instabilities~\cite{Baier:2000sb}. In the absence of general non-equilibrium formulation of QCD kinetic theory~\cite{Kurkela:2011ti}, we use QCD kinetic theory with isotropic approximations, which remove the unstable modes, to study the equilibration in expanding systems. Note that there are no unstable fermionic modes~\cite{Mrowczynski:2001az,Schenke:2006fz}.

 While the expansion conserves total energy, the local energy density in a given rapidity slice decreases as a function of time. At late times when the system is close to local thermal equilibrium, the time evolution of the temperature is given by ideal hydrodynamics with constant $T(\tau)\tau^{1/3}$. As the target temperature to which the out-of-equilibrium system aims to thermalize changes, so does the kinetic relaxation time. In the following we follow the practice of \cite{Kurkela:2018vqr} and, for each simulation, we determine the asymptotic value of $T(\tau)\tau^{1/3}|_{\tau \rightarrow \infty}$,  use the ideal hydrodynamics relation to extract what the temperature of the system would have been at earlier times if the full time evolution of the system were described by ideal fluid dynamics
\begin{align}
T_{\rm id.}(\tau) = \frac{(T(\tau)\tau^{1/3})|_{\tau \rightarrow \infty}}{\tau^{1/3}},\label{eq:Tid}
\end{align}
and use that in our definition of time dependent kinetic relaxation time
\begin{align}
\tau_R(\tau) = \frac{4\pi \eta/s}{T_{\rm id.}(\tau)}.\label{eq:tauRexp}
\end{align}

\begin{figure}
\centering
\includegraphics[width=0.9\linewidth]{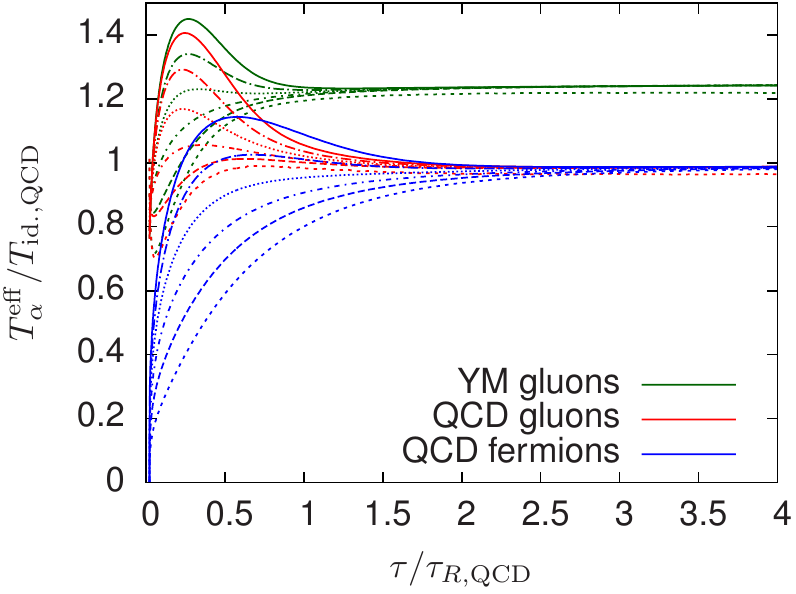}
\caption{
The effective temperature $T_\alpha^\text{eff}$ ($\alpha=-1,\ldots, 4$) evolution in longitudinally expanding system for QCD ($N_c=3$, $N_f=3$) and Yang-Mills ($N_c=3$, $N_f=0$).
Initial conditions given by  anisotropic over-occupied gluon state ($\lambda=5$, $\xi=10$, $\eta/s\approx 2.75$) with no initial fermions, \Eq{eq:focc}. Axes are scaled by time dependent relaxation time and asymptotic temperatures $\tau_R(\tau)=(4\pi \eta/s)/\TId(\tau)$ and $\TId(\tau)$.
}
\label{fig:plotteffprlgfexpl0}
\end{figure}

We consider an expanding system with over-occupied initial condition motivated by Color-Glass-Condensate framework
\begin{align}
f^g({\bf p}, \tau= \tau_0)=\frac{2 A}{\lambda} \frac{Q_0}{\sqrt{p_\perp^2+p_z^2\xi^2}} e^{-\frac{2}{3} \frac{p_\perp^2+\xi^2 p_z^2}{Q_0^2}},\label{eq:focc}
\end{align}
where the values of $A$ and $Q_0$, and $\xi$ are adjusted to reproduce the mean transverse momentum squared $\langle p_T^2 \rangle$ and energy density $e(\tau_0)$ at the initial time $\tau_0$ extracted from the classical lattice simulations of 
initial stages of the collision~\cite{Kurkela:2015qoa,Lappi:2011ju}. The anisotropy parameter  $\xi$ determining the ratio of longitudinal to transverse pressure is chosen such that $P_L \ll  P_T$ and is set in the following $\xi = 10$.  The same initial conditions have been studied in pure gauge theory in \cite{Kurkela:2015qoa, Keegan:2016cpi,Kurkela:2018vqr}. 
Here, as in \cite{Kurkela:2015qoa, Keegan:2016cpi,Kurkela:2018vqr}, we use $\tau_0 = 1/Q_s$, and $Q_0= 1.8 Q_s$ and we set $A=5.24$. Here $Q_s$ is the saturation scale of Color-Glass-Condensate, and is of order $Q_s^{-1} \sim 0.1 \rm fm$.

In \Fig{fig:plotteffprlgfexpl0} we show the time evolution of the effective temperatures $\Teff$ for the initial conditions \Eq{eq:focc} and $\lambda =5$, which corresponds to $\eta/s\approx 2.75$. The rapid longitudinal expansion quickly inverts the hierarchy of temperatures from an overoccupied state $f^g \gg 1$ ($T_\alpha^{\rm eff} > T_{\alpha+1}^{\rm eff}$) to an underoccupied state $f^g \ll 1$, 
which is well understood as the first stage of the \emph{bottom-up thermalization} \cite{Baier:2000sb}.
This transition takes place well before a substantial number of fermions are produced as can be seen from the good overlap of the time evolution of the effective temperatures in pure gauge theory with the full theory in  \Fig{fig:plotteffprlgfexpl0}.  
This suggests that during this early phase the presence of fermions does not significantly affect the evolution of bulk quantities. 
Note that the effective temperatures are sensitive only to the angular averaged distribution function, but because of the longitudinal expansion the system is highly anisotropic.

Next we look at the chemical equilibration of fermion energy fraction of the equilibrium energy density.
\begin{figure}
\centering
\includegraphics[width=0.9\linewidth]{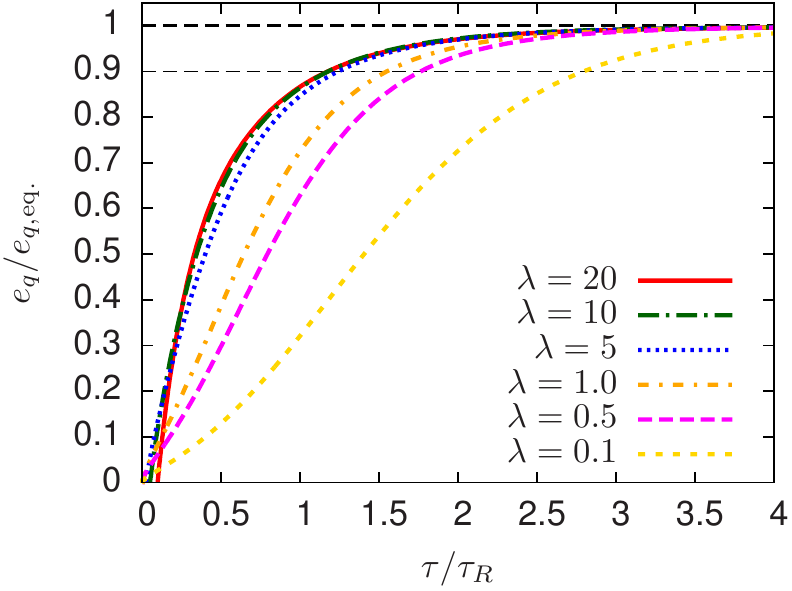}
\caption{Fermion energy fraction in a longitudinally expanding system with different coupling constants $\lambda$.
The system is initialized with an anisotropic over-occupied gluon state, \Eq{eq:focc} ($\xi=10$). The time axis is scaled by corresponding kinetic relaxation time $\tau_R(\tau)$, \Eq{eq:tauRexp} for each value of $\lambda$. 
}
\label{fig:prl_e_exp}
\end{figure}
In \Fig{fig:prl_e_exp} we show the approach to chemical equilibrium for different values of the coupling constant $\lambda$ and over-occupied initial state.  Using the same ad-hoc criterion of chemical equilibration as in non-expanding system \Eq{eq:chem}, we find that chemical equilibration happens for $\tau_\text{chem}\sim 1-2 \tau_R$ for a wide range of coupling constants $0.5\leq\lambda \leq 20$ and only for very small coupling  $\lambda =0.1$, we get $\tau_\text{chem} \approx 2.8 \tau_R$. We would like to remind that parametrically the chemical equilibration scales as $\sim (\lambda^2 T)^{-1}$, so in physical units the equilibration times change by several orders of magnitude. Scaling with relaxation time \Eq{eq:tauRexp}, reduces this vast separation of scales and for couplings $\lambda>1$ we observe the collapse of equilibration dynamics to the same curve and the chemical equilibration is reached at $\tau_\text{chem}\approx 1.2\tau_R$.
The difference in equilibration times for $\lambda\leq 1$, which was not present in the non-expanding case, arises from the additional scale --- the expansion rate $1/\tau$. Since the starting time $\tau_0=1/Q_s$ is kept fixed, for smaller values of the coupling constant, the system experiences a longer phase where the anisotropy is still increasing, i.e.\ the first stage of bottom-up, which delays equilibration.

\begin{figure}
\centering
\subfig{a}{\includegraphics[width=0.9\columnwidth]{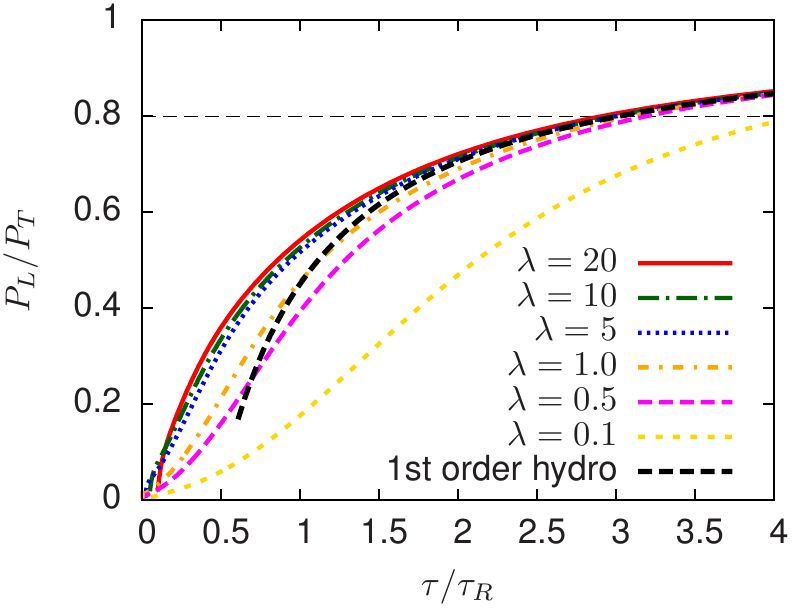}}
\subfig{b}{\includegraphics[width=0.9\columnwidth]{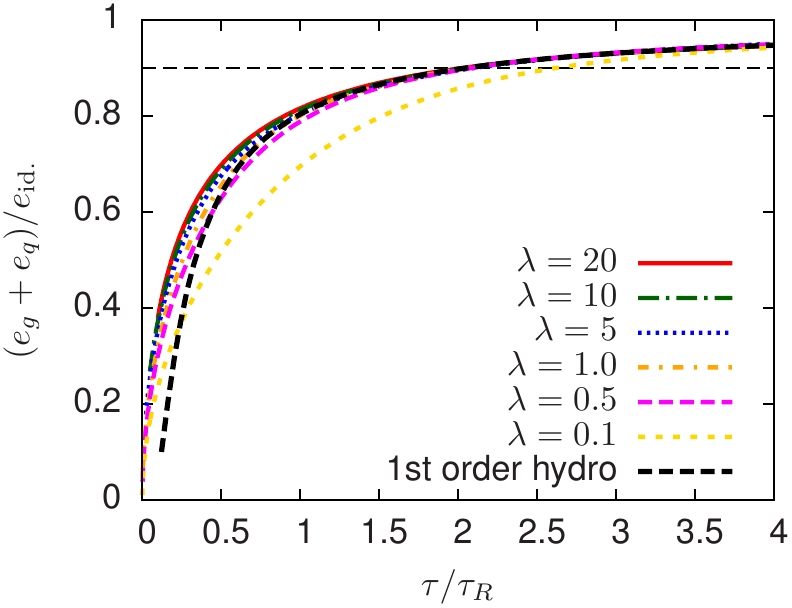}}
\caption{ (a) Total pressure anisotropy $P_L/P_T$ evolution in kinetic theory with over-occupied initial conditions \Eq{eq:focc} (b) Total energy density evolution relative to ideal estimate $e/e_\text{id.}=(T/\TId)^4$ given by \Eq{eq:Tid}. Comparison with first order viscous temperature \Eq{eq:T1st} shown by the black dashed curve.}
\label{fig:plotplptratioprlgfexpxt1}
\end{figure}

We note that the chemical equilibration takes place when the system is still highly anisotropic. This is seen in Fig.~\ref{fig:plotplptratioprlgfexpxt1}(a) depicting the time evolution of the ratio of the longitudinal and transverse pressures
\begin{align}
P_L &= \int \frac{d^3 p}{(2\pi)^3} \frac{p_z^2}{p^0} (\nu_g f_g+ 2 N_f\nu_q f_q),\\
P_T &= \frac{1}{2}\int \frac{d^3 p}{(2\pi)^3} \frac{p_x^2+p_y^2}{p^0} (\nu_g f_g+ 2 N_f \nu_q f_q).
\end{align}
The system becomes isotropic only at very late times. At the time of chemical equilibration ($\tau\lesssim 3 \tau_R$) the pressure anisotropy is still large $P_L/P_T<0.8$ for all values of $\lambda$.
However similarly to the pure gauge theory, the system's time evolution is well described by fluid dynamics well before pressure anisotropies become small; that is, the system exhibits ``hydrodynamization without thermalization.''

We quantify the approach to thermal equilibrium and hydrodynamization by defining two additional timescales $\tau_\text{therm}$ and $\tau_\text{hydro}$, in analogy to \Eq{eq:chem}. We require the combined gluon and fermion energy density $e=e_g+e_q$ to be within 10\% of ideal and viscous hydrodynamic estimates of energy density
\begin{align} 
\frac{e(\tau_\text{therm})}{e_{{\rm id.}}} = \left(\frac{T(\tau_\text{therm})}{T_{\rm id}(\tau_\text{therm})}\right)^4= 0.9 \label{eq:therm},\\
\left|1-\frac{e(\tau_\text{hydro})}{e_{{\rm 1st}}}\right| =\left|1- \left(\frac{T(\tau_\text{hydro})}{T_{\rm 1st}(\tau_\text{hydro})}\right)^4\right|= 0.1 \label{eq:hydro}\,.
\end{align}
Here $T_\text{id}$ is the ideal estimate of the local temperature \Eq{eq:Tid} and
 $T_\text{1st}$ is the 1st order viscous hydrodynamic solution of longitudinally expanding system with the same late time asymptotics as $T_\text{id}$. The analytical solution for first order conformal hydrodynamical equations of motion can be written in units of $\tau_R$ as~\cite{Kouno:1989ps,Muronga:2001zk}
\begin{equation}
\frac{T_\text{1st}(\tau)}{T_\text{id}(\tau)}=1-\frac{2}{12\pi}\frac{\tau_R}{\tau}.\label{eq:T1st}
\end{equation}
In \Fig{fig:plotplptratioprlgfexpxt1}(b) we show the time evolution of the total energy density scaled by ideal estimate, i.e.\ $e/e_\text{id.}= (T/\TId)^4$. We find that 90\% of equilibrium energy is reached at $\tau_\text{therm}\sim 2 \tau_R$ for the coupling constant values $0.5 \leq\lambda \leq 20$. Only for $\lambda =0.1$ the approach to equilibrium is slower and this criterion is satisfied at $\tau_\text{therm}\sim 2.5$. 
Next, we use  the 1st order hydrodynamic estimate for temperature \Eq{eq:T1st} and compare it to the kinetic theory\footnote{Note that substituting  \Eq{eq:T1st} in \Eq{eq:hydro} generates terms, which are formally higher in viscous gradients and could be dropped at first order.  We do not do such expansion in \Fig{fig:plotplptratioprlgfexpxt1} and use the full temperature estimate \Eq{eq:T1st}.}. We achieve agreement with full kinetic theory evolution at very early times and by the time $\tau_\text{hydro}\lesssim 0.5 \tau_R$  the criterion \Eq{eq:hydro} is satisfied for $0.5 \leq\lambda \leq 20$. For $\lambda=0.1$  thus defined hydrodynamization takes place somewhat later at $\tau\sim 1.3\tau_R$.

\section{Conclusions\label{sec:conclusions}}

In this paper we presented a complete simulation of chemical equilibration in leading order QCD kinetic theory in stationary and expanding systems with infinite transverse extent. By analysing how out-of-equilibrium plasma of $N_c=3$ gluons and $N_f=3$ quarks relax to the common thermal equilibrium for different values of the coupling constant $\lambda$, we determined the chemical  equilibration time in non-expanding isotropic systems, which we define by requiring the quark energy fraction to be within 10\% of their equilibrium value. For initial conditions with no quarks present, thus defined chemical equilibrium is reached at time $t_\text{chem}\sim 1.1{-}1.4\tau_R$, where $\tau_R=(4\pi\eta/s)/T_\text{final}$ is the kinetic relaxation time and $\eta/s(\lambda=10,1.0,0.1)\approx 1,35,1900$.
 We also note faster gluon dynamics, which results in gauge bosons reaching kinetic equilibrium among themselves before thermalizing with fermions. Consequently, for the case of the over-occupied gluon initial state, gluons first  equilibrate through a self-similar cascade as in pure glue theory at times $t_\text{kinetic}^g\sim 0.2 \tau_R$, and then the equilibrium quark densities are produced by the quasi-thermal gluon background. We would like to emphasise that the chemical equilibration time dependence on the coupling constant $\lambda$---the only microscopic parameter of the theory ---is very well captured by the specific shear viscosity $\eta/s$, which is the macroscopic QGP property. As the relaxation time $\tau_R$ changes by orders of magnitude, the equilibration dynamics in rescaled units $t/\tau_R$ remains largely unchanged.

Next we  studied the QCD equilibration in homogeneous, but longitudinal expanding systems,  which is a relevant situation for heavy ion phenomenology. There the expansion prevents the system from ever reaching static thermal equilibrium and one instead may define thermalization  time $\tau_\text{therm}$ by requiring the total energy density to be within 10\% of the value given by ideal hydrodynamic evolution $e_\text{id.}(\tau)=(e\tau^{4/3})_\infty \tau^{-4/3}$.  Here we note that in the expanding case the effective temperature of the plasma is decreasing and so we defined \emph{time-dependent}  kinetic relaxation time $\tau_R(\tau)\sim \tau^{1/3}$ which grows in time, see \Eq{eq:tauRexp}. 
For Color-Glass-Condensate motivated, anisotropic and overoccupied initial conditions, \Eq{eq:focc}, such thermalization happens at $\tau_\text{therm}\sim 2\tau_R(\tau)$ for a range of the coupling constants $0.5\leq \lambda \leq 20$, but is somewhat delayed for $\lambda=0.1$.
 The chemical composition of the plasma changes rapidly in the first couple units of relaxation time. Similarly to the non-expanding systems, gluons undergo a kinetic equilibration faster than fermions, in agreement with two-stage QGP equilibration argued in Ref.~\cite{Shuryak:1992wc}. However only when the expansion rate slows down  and viscous corrections to the particle distribution function are small enough, particle distributions are well approximated by the equilibrium Bose-Einstein or Fermi-Dirac distributions. 
Keeping in mind that the effective kinetic relaxation time $\tau_R$ is growing in time as the temperature is decreasing, the chemical equilibration in longitudinally expanding systems for moderate values of the coupling constants $\lambda=5,10,20$ ($\alpha_s\sim 0.1{-}0.5$) proceeds very similarly in rescaled units to the non-expanding case and chemical equilibration is reached at $\tau_\text{chem}\sim 1.2\tau_R(\tau)$.
For smaller values of the coupling constant $\lambda\leq 1$ we do not see the collapse to the same universal curve and the  chemical equilibration (in units of $\tau_R$) takes place later.

In summary, 
the chemical composition is an important property of the expanding QGP fireball, which is not captured by conventional hydrodynamic modelling of heavy ion collisions, but is essential for the hadrochemistry, photon production, and determines which equation of state best describes the medium. It would be therefore interesting to study if a generalization of hydrodynamics involving nearly conserved charges could be used to describe this non-equilibrium evolution.
For realistic values of the coupling constant $\alpha_s\sim 0.3$, we find that even in expanding systems the coupling constant dependence can be factored out by  rescaling  time with kinetic relaxation time $\tau_R=(4\pi \eta/s)/T_\text{id.}$, which results in the following ordering of hydrodynamization, chemical equilibration and thermalization timescales
\begin{equation}
\underbrace{\tau_\text{hydro}}_{\lesssim 0.5\tau_R}<\underbrace{\tau_\text{chem}}_{\sim 1.2\tau_R}<\underbrace{\tau_\text{therm}}_{\sim 2\tau_R},
\end{equation}
according to criteria given in \Eqs{eq:chem}, \eq{eq:hydro} and \eq{eq:therm}.
Such universality allows one to convert the dimensionless time $\tau/\tau_R$ to physical units by matching the late time constants $(\tau^{1/3} T)_\infty$ and $\eta/s$ from hydrodynamical modelling of heavy ion collisions and which is explored in our companion paper~\cite{Kurkela:2018xxd}.

\noindent \textbf{Acknowledgements:} 
The authors thank  Peter Arnold, J\"urgen Berges, Ulrich Heinz, Jacopo Ghiglieri, Jean-Fran\c cois Paquet, S\"oren Schlichting, Derek Teaney, and Urs Wiedemann for valuable discussions.
This work was supported in part by the German Research Foundation (DFG) 
Collaborative Research Centre (SFB) 1225 (ISOQUANT) (A.M.). Finally, A.M.
thanks CERN Theoretical Physics
Department for the hospitality during the short-term visit.

\appendix

\section{Specific shear viscosity in QCD kinetic theory\label{sec:etas}}
The dynamical simulations of the Boltzmann equation with leading order QCD kinetic theory collisions kernels, \Eq{eq:bolz}, allows for the direct determination of the QGP transport properties, for example, the shear viscosity over entropy ratio $\eta/s$. The transport coefficients extracted this way can be compared to calculations of  $\eta/s$ using  diagonalization of (linearized) collision kernels around thermal equilibrium~\cite{Arnold:2003zc}. In \Fig{fig:plotetasvslambda} we show the specific shear viscosity as a function of the coupling constant $\lambda$ obtained from the  effective kinetic theory  simulations with different, but leading order equivalent IR regulators of the elastic scattering matrix element \Eq{eq:2to2}. The first regulator corresponds to the scheme given by \Eq{eq:reg}, while in the second case we insert additional $m^2_s/(q^2+m^2_s) \sim \mathcal{O} (g^2)$ factors to guarantee the positivity of the scatter matrix element $|\mathcal{M}^{ab}_{cd}|^2$. For  $\lambda \lesssim 2$ different implementations of the kinetic theory agree with each other at $\sim 10\%$. The same level of agreement is also seen with the next-to-leading-log formula, which is a good approximation for the full leading order results~\cite{Arnold:2003zc} (and which corresponds to yet another IR completion of the kinetic theory).
For completeness below we summarize the extracted values of $\eta/s$ used in the paper
\begin{center}
\begin{tabular}{|c|c|c|c|c|c|c|}
\hline 
$\lambda$ &0.1 & 0.5  & 1.0 & 5.0 & 10  & 20  \\ 
\hline 
$\eta/s$ &1900 & 114 & 35 & 2.75 & 1.0 & 0.39  \\ 
\hline 
\end{tabular} 
\end{center}

\begin{figure}
\centering
\includegraphics[width=0.9\linewidth]{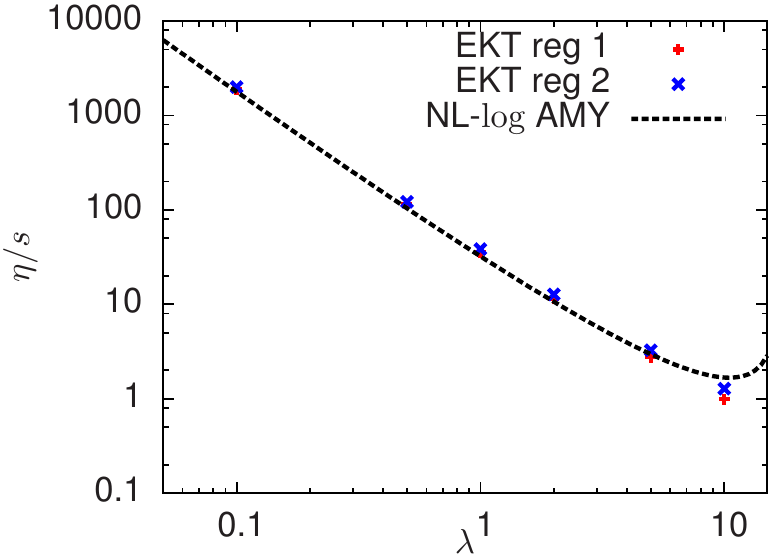}
\caption{Shear viscosity over entropy ratio as a function of the coupling constant $\lambda=N_c g^2$ in our leading order kinetic theory implementation with two regularization schemes of the elastic collision kernel (see the text). The dashed line corresponds to the next-to-leading-log result from Ref.~\cite{Arnold:2003zc}. }
\label{fig:plotetasvslambda}
\end{figure}

\bibliography{master.bib}

\end{document}